\documentclass[aip,reprint, jmp, amsmath, amssymb]{revtex4-1}
\usepackage{graphicx}% Include figure files
\usepackage{dcolumn}% Align table columns on decimal point
\usepackage{amsmath}
\usepackage{bm}% bold math
\usepackage{txfonts}
\begin{document}

\title{Realization of tunable spin-dependent splitting in intrinsic photonic spin Hall effect}% Force line breaks with \\

\author{Xiaohui Ling}
\affiliation{SZU-NUS Collaborative Innovation Center for
Optoelectronic Science and Technology, and Key Laboratory of
Optoelectronic Devices and Systems of Ministry of Education and
Guangdong Province, College of Optoelectronic Engineering, Shenzhen
University, Shenzhen 518060, China} \affiliation{Laboratory for spin
photonics, College of Physics and Microelectronic Science, Hunan
University, Changsha 410082, China} \affiliation{Department of
Physics and Electronic Information Science, Hengyang Normal
University, Hengyang 421002, China}

\author{Xunong Yi}
\affiliation{SZU-NUS Collaborative Innovation Center for
Optoelectronic Science and Technology, and Key Laboratory of
Optoelectronic Devices and Systems of Ministry of Education and
Guangdong Province, College of Optoelectronic Engineering, Shenzhen
University, Shenzhen 518060, China}

\author{Xinxing Zhou}
\affiliation{Laboratory for spin photonics, College of Physics and
Microelectronic Science, Hunan University, Changsha 410082, China}

\author{Yachao Liu}
\affiliation{Laboratory for spin photonics, College of Physics and
Microelectronic Science, Hunan University, Changsha 410082, China}

\author{Weixing Shu}
\affiliation{Laboratory for spin photonics, College of Physics and
Microelectronic Science, Hunan University, Changsha 410082, China}

\author{Hailu Luo}\email{hailuluo@hnu.edu.cn}
\affiliation{SZU-NUS Collaborative Innovation Center for
Optoelectronic Science and Technology, and Key Laboratory of
Optoelectronic Devices and Systems of Ministry of Education and
Guangdong Province, College of Optoelectronic Engineering, Shenzhen
University, Shenzhen 518060, China} \affiliation{Laboratory for spin
photonics, College of Physics and Microelectronic Science, Hunan
University, Changsha 410082, China}

\author{Shuangchun Wen}
\affiliation{Laboratory for spin photonics, College of Physics and
Microelectronic Science, Hunan University, Changsha 410082, China}

\date{\today}% It is always \today, today,

\begin{abstract}
We report the realization of tunable spin-dependent splitting in
intrinsic photonic spin Hall effect. By breaking the rotational
symmetry of a cylindrical vector beam, the intrinsic vortex phases
that the two spin components of the vector beam carries, which is
similar to the geometric Pancharatnam-Berry phase, is no longer
continuous in the azimuthal direction, and leads to observation of
spin accumulation at the opposite edge of the beam. Due to the
inherent nature of the phase and independency of light-matter
interaction, the observed photonic spin Hall effect is intrinsic.
Modulating the topological charge of the vector beam, the
spin-dependent splitting can be enhanced and the direction of spin
accumulation is switchable. Our findings may provide a possible
route for generation and manipulation of spin-polarized photons, and
enables spin-based photonics applications.
\end{abstract}

\maketitle Photonic spin Hall effect (SHE) describes the mutual
influence of the photon spin (polarization) and the trajectory
(orbit angular momentum) of light beam propagation, i.e., spin-orbit
interaction~\cite{Onoda2004,Bliokh2006,Bliokh2008A}. It manifests as
spin-dependent splitting (SDS) of light, which corresponds to two
types of geometric phases: the Rytov-Vladimirskii-Berry phase
associated with the evolution of the propagation direction of light
and the Pancharatnam-Berry phase related to the manipulation with
the polarization state of
light~\cite{Bliokh2008A,Bliokh2008B,Hosten2008}. When a light beam
reflecting/refracting at a planar interface or passing through an
inhomogeneous anisotropic medium, it may acquire a locally varying
geometric phase, i.e., the different part of the beam carrying
different geometric
phase~\cite{Hosten2008,Qin2009,Bomzon2001,Marrucci2006,Ren2012}. The
interference upon transmission leads to the redistribution of the
beam intensity and may show a SDS of light, that is, the photonic
SHE. Recent advances in this field provide new opportunities for
advantageous measurement of the optical parameters of nanostructures
such as metallic film and graphene~\cite{Zhou2012A,Zhou2012B}. More
importantly, it offers a possible way for generation and
manipulation of spin-polarized photons and spin/orbital angular
momenta of light, and enables spin-controlled photonics
applications~\cite{Shitrit2013,Yin2013}.

At the interface reflection and refraction of different media, the
SDS induced by the photonic SHE is generally very tiny and sensitive
to the optical parameters of the media (e.g., refractive indices and
thickness, etc.) which makes it difficult to manipulate the SDS with
a real medium interface~\cite{Hosten2008,Qin2009,Luo2011}. An
inhomogeneous anisotropic medium can produce a giant and tunable SDS
in momentum space~\cite{Shitrit2011,Ling2012,Li2013}, but it
requires a complex and precise fabrication technique to construct
the medium. Actually, the photonic SHE does not always rely on the
light-matter interaction; it can be observed in an oblique
observation plane respect to the beam propagation direction even in
the free space~\cite{Aiello2009,Korger2014}. This effect is
intrinsically dependent upon the polarization geometry of the beam
projected on the oblique observation plane rather than any kind of
light-matter interaction. Similar to the photonic SHE occurring at
the interface reflection and refraction, the induced SDS is fixed
and also exceedingly weak.

In this work, we report the realization of tunable SDS in intrinsic
photonic SHE by blocking part of a cylindrical vector beam (CVB)
with a fan-shaped aperture (FSA). The underlying mechanism is
attributed to the inherent, opposite vortex phase that the two spin
components (circular polarizations) of the CVB carry, so the
observed photonic SHE is intrinsic. This phase is similar to the
geometric Pancharatnam-Berry phase which creates a geometric phase
gradient in momentum space, and results in the SDS~\cite{Ling2014}.
By modulating the topological charge of the CVB, the SDS can be
enhanced, and the direction of spin accumulation is switchable.

\begin{figure}
\centerline{\includegraphics[width=8cm]{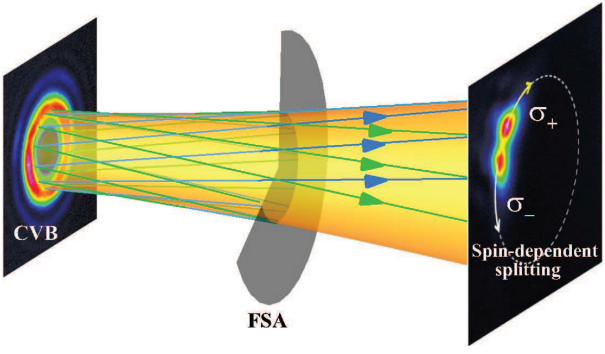}}
\caption{\label{Fig1} Schematic illustration of the intrinsic
photonic SHE of the CVB. The blue and green rays with arrows
indicate the local wave vectors of the two spin components of the
CVB. The rotational symmetry is broken by blocking part of the CVB
with a FSA. It shows a direct intensity splitting of the left- and
right-handed circular polarization components ($\sigma_+$ and
$\sigma_-$). }
\end{figure}

The CVB that exhibits inhomogeneous polarization distribution with
rotational symmetry has drawn great attention due to its great
potential in many fields including optical manipulation, nonlinear
optics, and optical communications (see ~\cite{Zhan2009} for a
review and the references therein). It can be viewed as
superposition of two sub-beams carrying opposite spin angular
momentum (circular polarization) and opposite orbital angular
momentum (vortex phase), and can be geometrically represented by the
so-called higher order Poincar\'{e}
sphere~\cite{Holleczek2011,Milione2011}. The algebraical description
is represented by the following equation:
\begin{equation}
|\psi\rangle=\cos\left(\frac{\phi}{2}\right) |R\rangle
e^{-i\beta}+\sin\left(\frac{\phi}{2}\right) |L\rangle
e^{i\beta}.\label{dec}
\end{equation}
Here, $\phi$ is a tuning parameter. $|R\rangle e^{i\beta}$ and
$|L\rangle e^{-i\beta}$ are circularly polarized vortex light, with
$|R\rangle$ and $|L\rangle$ standing for the right- and
left-circular polarizations, respectively. For $\phi=\pi/2$,
Eq.~(\ref{Fig1}) indicates a linear polarized CVB. In this case, the
Jones vector of the CVB can be simply written as ($\cos\beta,
\sin\beta)^T$ where $\beta=m\varphi+\beta_0$ with $m$ the
topological charge, $\varphi$ the azimuthal angle, and $\beta_0$ a
constant. Other values of $\phi$ represent elliptical polarized CVB.
Equation~(\ref{Fig1}) unambiguously illustrates that the two
circular polarizations carry just opposite azimuthal vortex phase
$e^{\pm i\beta}$. This phase is similar to the geometric
Pancharatnam-Berry phase which can be obtained in some inhomogeneous
anisotropic
media\cite{Bomzon2001,Marrucci2006,Niv2008,Ling2012,Karimi2014}.
Although the two components have opposite vortex phases and local
energy flows, their superposition does not show a helical wave
front. They always superpose exactly at the same position and no SDS
can be observed, due to the rotational symmetry. When blocking part
of a vortex, its intensity distribution and the geometric shadow
area just behind the obstacle rotate in the sense of the vortex's
handedness~\cite{Arlt2003,Davis2005}, so the two spin components of
the CVB no longer superpose exactly and separate from each other
(see the schematic picture in Fig.~\ref{Fig1}).

\begin{figure}
\centerline{\includegraphics[width=8cm]{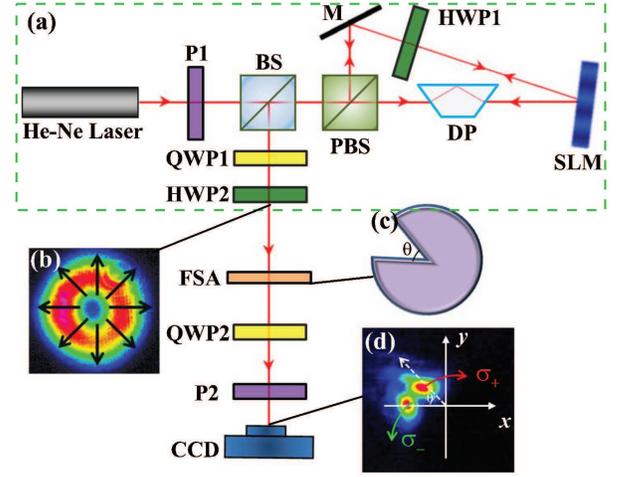}}
\caption{\label{Fig2} Experimental setup for generating the CVB and
measuring the photonic SHE when it passing through a fan-shaped
aperture (FSA). (a) The laser source is a single mode linearly
polarized He-Ne laser, wavelength $\lambda=632.8$ nm. The polarizer
(P1) ensures a $45^\circ$ polarization light to impinge into the
polarization beam splitter (PBS) and being equally split into two
sub-beams. The Sagnac interferometer comprised of a PBS, a
phase-only spatial light modulator (SLM, Holoeye Pluto-Vis,
Germany), a Dove prism (DP), a half-wave plate (HWP1) and a mirror
(M). A quarter-wave plate (QWP1) changes the two sub-beams from
linear polarization to opposite circular polarizations. Then a CVB
is produced after the QWP1. Another half-wave plate (HWP2) is used
to modulate the polarization distribution of the CVB. BS represents
a non-polarizing beam splitter. (b) An example of the generated CVB
with radial polarization distribution. (c) Schematic picture of the
FSA. (d) CCD recorded intensity of the CVB passing through the FSA
without the Stokes parameter measurement setup (QWP2 and P2). It
shows a direct intensity splitting of the $\sigma_+$ and $\sigma_-$
components.}
\end{figure}

To measure the intrinsic photonic SHE, we set up a Sagnac
interferometer to generate the linear polarized CVB, as shown in
Fig.~\ref{Fig2}(a), which can also be conveniently generated by many
other methods~\cite{Zhan2009,Bomzon2002,Wang2007,Jones2009}. This
apparatus relies on the superposition of two equal-intensity beams
with opposite circular polarizations and opposite vortex phases,
according to Eq.~(\ref{dec}). The polarizer (P1) can ensure the
light output from the He-Ne laser to be $45^\circ$ polarization
respect to the horizontal direction. Then the beam passes through
the polarization beam splitter (PBS) and is split into two
equal-intensity beams with the transmission beam being horizontal
polarization and the reflection beam vertical polarization. The two
sub-beams propagate exactly in a common path. A phase-only spatial
light modulator (SLM) is used at small incidence angle, and can
apply a vortex phase with any desired topological charge to a
horizontal polarization beam which is a good approximation of phase
vortex-bearing Laguerre-Gauss beam. A half-wave plate (HWP1) with
its optical axis $45^\circ$ inclined to the horizontal direction is
employed to change the vertical polarization to a horizontal one and
vice versa for its counter-propagating counterpart. A Dove prism
(DP) involves one reflection to change the sign of the topological
charge alone one beam path, and ensures that the output beam
contains two opposite phase vortexes. Then we use a quarter-wave
plate (QWP1) with $45^\circ$ optical axis orientation changes the
two sub-beams into opposite circular polarizations. So the CVB is
generated after the QWP1 [see Fig.~\ref{Fig2}(b)] and its intensity
shows a donut-shaped profile similar to a vortex beam. The HWP2 can
help to modulate the polarization distribution of the CVB, e.g.,
changing a radial polarization into an azimuthal polarization or any
intermediate states~\cite{Ling2014}.

The generated CVB then passes through a FSA [Fig.~\ref{Fig2}(c)] at
normal incidence. For the sake of simplicity and without loss of
generality, the FSA can be described by the following expression:
\begin{equation} \label{eq:1}
T(\theta)=\left\{ \begin{aligned}
1~~\text{for}~\theta~\text{radian}  \\
0~~~~\text{otherwise}.
\end{aligned} \right.
\end{equation}

It is known that the Stokes parameter $S_3$ can be used to describe
the circular polarization degree~\cite{Born1999}, so the SDS of
light can be obtained by measuring the $S_3$ parameter pixel by
pixel in the output using a typical setup: a quarter-wave plate
(QWP2), a polarizer (P2), and a CCD camera. In the experiments, the
$S_3$ parameter can be given by
\begin{equation}
S_3=\frac{I_{\sigma+}-I_{\sigma-}}{I_{\sigma+}+I_{\sigma-}}.
\end{equation}
Here, $I_{\sigma+}$ and $I_{\sigma-}$ represent the intensities
measured in the circular polarization basis, respectively.

We first consider the influence of the topological charge $m$ on the
intrinsic photonic SHE. The vortex phase creates a phase gradient in
the azimuthal direction, which results in a SDS in $k$ (momentum)
space: $\Delta k=\sigma_{\pm}\nabla\beta=\sigma_{\pm}m
\hat{e}_{\varphi}$ with $\sigma_{\pm}=\pm 1$ representing the left
and right circular polarization and $\hat{e}_{\varphi}$ the unit
vector in the azimuthal direction,
respectively~\cite{Shitrit2013,Ling2014}. Hence, this shift is
proportional to the value of $m$. However, for a CVB, the SDS cannot
be observed in free-space propagation, due to its rotational
symmetry. Breaking the rotational symmetry, it is expected to
observe the spin accumulation at the edge of the beam.

\begin{figure}
\centerline{\includegraphics[width=8.5cm]{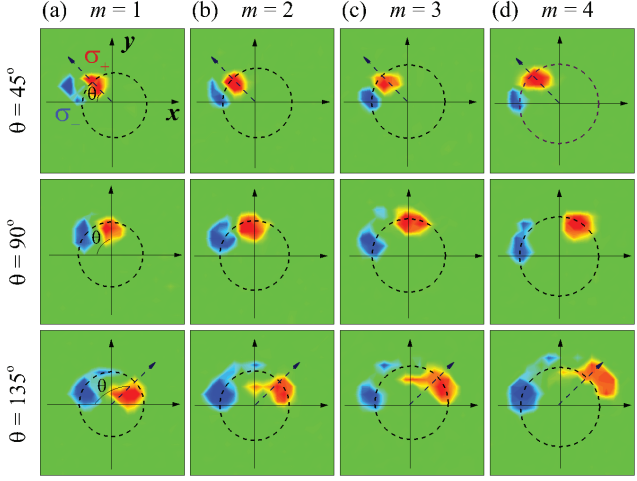}}
\caption{\label{Fig3} Intrinsic photonic SHE of the CVB with
different topological charges for different FSAs with three typical
aperture angles $\theta$. The dashed circles indicate the profile of
the incident CVB. The experimentally measured light spot of the
$S_3$ parameters have a little deviation from the dashed circles
because of the unavoidable experimental errors in the process of the
$S_3$ measurement.}
\end{figure}

\begin{figure}[b]
\centerline{\includegraphics[width=8.5cm]{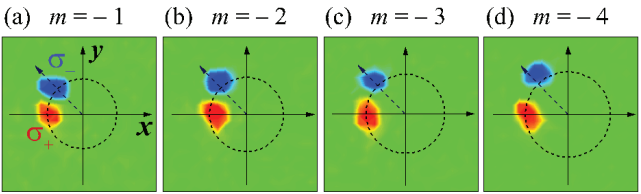}}
\caption{\label{Fig4} Intrinsic photonic SHE of the CVB with
negative topological charges $m=-1, -2, -3$, and $-4$ for the
aperture angle of the FSA $\theta=45^\circ$.}
\end{figure}

Figure~\ref{Fig3} shows the measured $S_3$ parameters of the
photonic SHE for different CVBs and different aperture angles
$\theta$. The spin-dependent shift increases with the rise of the
value of $m$. On the other hand, the shift distance is limited by
the dimension of the aperture angle because the spin-polarized
photons accumulate at the beam edge. Also because of this, the
spin-dependent splitting increases with the increase of $\theta$. If
reversing the sign of $m$ by modulating the phase picture displayed
on the SLM, the direction of spin accumulation is also inverted, as
shown in Fig.~\ref{Fig4}. Because the sign of $m$ directly
determines the handedness of the vortex phase that the two spin
components of the CVB carry. The measured $S_3$ parameters have a
little deviation from the expected position (dashed circles in
Figs.~\ref{Fig3} and~\ref{Fig4}) due to the unavoidable experiment
errors in the $S_3$ measurement.

\begin{figure}
\centerline{\includegraphics[width=7.5cm]{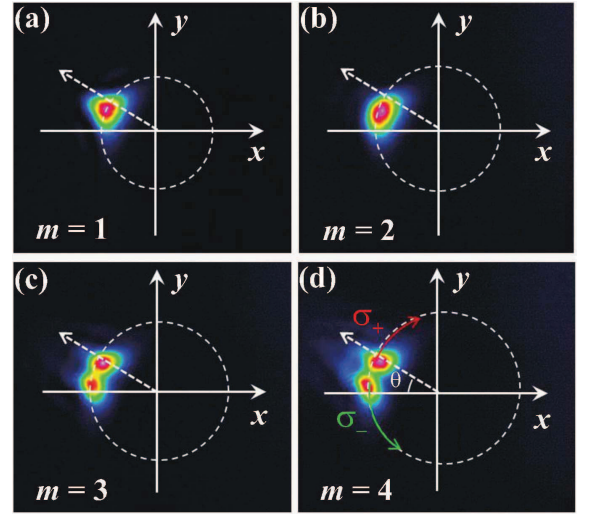}}
\caption{\label{Fig5}  Direct intensity illustration of the giant
photonic SHE with the aperture angle of the FSA $\theta=30^\circ$.
The four experimental screenshots show the results of CVBs with
different topological charges $m$.}
\end{figure}

The SDS of the intrinsic photonic SHE of the CVB can be large enough
for direct measurement without using the weak measurement
technology~\cite{Hosten2008,Dressel2014}. With the increase of the
value $m$ of the CVB, we can directly observe the intensity
separation of the $\sigma_+$ and $\sigma_-$ components, as shown in
Fig.~\ref{Fig5} for $\theta=30^\circ$. In (a) and (b) for $m=1$ and
2, the two components do not separate enough from each other and
show a single-spot profile, however, it still can be discriminated
by measuring the $S_3$ parameter just like in Figs.~\ref{Fig3}
and~\ref{Fig4}. If the phase gradient is large enough, the
$\sigma_+$ and $\sigma_-$ components are almost completely
separated, as shown in Fig.~\ref{Fig5}(c) and 5(d). The induced
spin-dependent shift is within millimeters (the beam waist of the
He-Ne laser is 0.7 mm and expanded to 2.1 mm by a beam expander),
which is many times larger than the optical wavelength (632.8 nm).
It is also much larger than that had observed previously in beam
reflection and refraction with the shift of the order of a fraction
of wavelength~\cite{Hosten2008,Qin2009,Yin2013}. This enables us to
observe a giant photonic SHE.

Actually, the focusing behavior of the CVB with axial symmetry
broken by a lens has been explored, and its application in optical
trapping was suggested~\cite{Wang2011}. In the focal plane, the
circular polarizations have a spin-dependent rotation with their
rotation angle reaching to $\pi$/2 relative to the aperture
edge~\cite{Arlt2003}. This is due to that the scalar vortex beam has
an azimuthal energy flow along the circumference of the beam when it
propagates~\cite{Padgett1995}. In our context, we consider the
intrinsic SHE of the CVB with rotational symmetry breaking at a
propagation distance much less than the Rayleigh distance, so the
spin-polarized photons accumulate at the opposite edge of the beam.
As mentioned above, the SDS occurs in the $k$ space in the azimuthal
direction, the induced shift would increase linearly upon beam
propagation. Opposite topological charge just reverses the direction
of spin accumulation.

In summary, we have experimentally demonstrated the realization of
tunable SDS in intrinsic photonic SHE of the CVB by breaking its
rotational symmetry using a FSA to block part of the CVB. The spin
accumulation occurs at the edge of the beam, and the SDS increases
with the topological charge of the CVB and restricts by the aperture
angle of the FSA. The underlying mechanism is attributed to the
discontinuous local energy flow that results from the broken,
opposite vortex phases. It is large enough to be directly observed
without using a weak measurement technology. Because of the inherent
nature of the phase and independency of light-matter interaction,
the observed photonic SHE is intrinsic. This enables us to observe a
direct and giant photonic SHE. Our findings reveal that the photonic
SHE may be manipulated (enhanced or inverted) by directly tailoring
the polarization geometry of light, which may provide a possible
route for generation and manipulation of spin-polarized photons, and
enables spin-controlled photonics applications.

This research was supported by the National Natural Science
Foundation of China (Grants No. 61025024, No. 11274106, and No.
11347120), the Scientific Research Fund of Hunan Provincial
Education Department of China (Grant No. 13B003), and the Doctorial
Start-up Fund of Hengyang Normal University (Grant No. 13B42).

\end{document}